# Radiation Tests of Real-Sized Prototype RPCs for the Phase-2 Upgrade of the CMS Muon System

K. S. Lee[*,a], S. Choi, B. S. Hong, M. Jo, J. W. Kang, M. Kang, H. Kim, K. Lee, S. K. Park[a], A. Cimmino, S. Crucy, A. Fagot, M. Gul, A. A. O. Rios, M. Tytgat, N. Zaganidis[b], S. Ali, Y. Assran, A. Radi, A. Sayed[c], G. Singh[d], M. Abbrescia, G. Iaselli, M. Maggi, G. Pugliese, P. Verwilligen[e], W. van Doninck[f], S. Colafranceschi, A. Sharma[g], L. Benussi, S. Bianco, D. Piccolo, F. Primavera[h], V. Bhatnagar, R. Kumar, A. Metha, J. Singh[i], A. Ahmad, M. Ahmad, Q. Hassan, H. R. Hoorani, W. A. Khan, T. Khurshid[j], D. H. Kim[k], S. K. Nam[l], M. Goutzvitz, G. Grenier, F. Lagarde, I. B. Laktineh[m], S. Carpinteyro Bernardino, C. Uribe Estrada, I. Pedraza[n], S. Carrillo Moreno, F. Vazquez Valencia[o], L. M. Pant[p], S. Buontempo, N. Cavallo, M. Esposito, F. Fabozzi, G. Lanza, L. Lista, S. Meola, M. Merola, I. Orso, P. Paolucci, F. Thyssen[q], A. Braghieri, A. Magnani, P. Montagna, C. Riccardi, P. Salvini, I. Vai, P. Vitulo[r], Y. Ban, S. J. Qian[s], M. Choi[t], Y. Choi, J. Goh, D. Kim[u], A. Aleksandrov, R. Hadjiiska, P. Iaydjiev, M. Rodozov, S. Stoykova, G. Sultanov, M. Vutova[v], A. Dimitrov, L. Litov, B. Pavlov, P. Petkov[w], D. Lomidze[x], C. Avila, A. Cabrera, J. C. Sanabriay[y], I. Crotty[z], J. Vaitkus[aa]

[a]*Korea University, Department of Physics, 145 Anam-ro, Seongbuk-gu, Seoul 02841, Republic of Korea*
[b]*Ghent university, Dept. of Physics and Astronomy, Proeftuinstraat 86, B-9000 Ghent, Belgium*
[c]*Egyptian Network for High Energy Physics, Academy of Scientific Research and Technology, 101 Kasr El-Einy St. Cairo Egypt*
[d]*Chulalongkorn University, Department of Physics, Faculty of Science, Payathai Road, Phatumwan, Bangkok, THAILAND - 10330*
[e]*INFN, Sezione di Bari, Via Orabona 4, IT-70126 Bari, Italy*
[f]*Vrije Universiteit Brussel, Boulevard de la Plaine 2, 1050 Ixelles, Belgium*
[g]*Physics Department CERN, CH-1211 Geneva 23, Switzerland*
[h]*INFN, Laboratori Nazionali di Frascati (LNF), Via Enrico Fermi 40, IT-00044 Frascati, Italy*
[i]*Department of Physics, Panjab University, Chandigarh Mandir 160 014, India*
[j]*National Centre for Physics, Quaid-i-Azam University, Islamabad, Pakistan*
[k]*Kyungpook National University, 80 Daehak-ro, Buk-gu, Daegu 41566, Republic of Korea*
[l]*Kangwon National University, 1 Kangwondaehak-gil, Chunchon 24341, Republic of Korea*
[m]*Universite de Lyon, Universite Claude Bernard Lyon 1, CNRS-IN2P3, Institut de Physique Nucleaire de Lyon, Villeurbanne, France*
[n]*Benemerita Universidad Autonoma de Puebla, Puebla, Mexico*
[o]*Universidad Iberoamericana, Mexico City, Mexico*
[p]*Nuclear Physics Division Bhabha Atomic Research Centre Mumbai 400 085, INDIA*
[q]*INFN, Sezione di Napoli, Complesso Univ. Monte S. Angelo, Via Cintia, IT-80126 Napoli, Italy*

---

[*] Corresponding author.

[r]*INFN, Sezione di Pavia, Via Bassi 6, IT-Pavia, Italy*
[s]*School of Physics, Peking University, Beijing 100871, China*
[t]*University of Seoul, 163 Seoulsiripdae-ro, Dongdaemun-gu, Seoul 02504, Republic of Korea*
[u]*Sungkyunkwan University, 2066 Seobu-ro, Jangan-gu, Suwon-si 16419, Gyeonggi-do, Republic of Korea*
[v]*Bulgarian Academy of Sciences, Inst. for Nucl. Res. and Nucl. Energy, Tzarigradsko shaussee Boulevard 72, BG-1784 Sofia, Bulgaria*
[w]*Faculty of Physics, University of Sofia,5, James Bourchier Boulevard, BG-1164 Sofia, Bulgaria*
[x]*Tbilisi University, 1 Ilia Chavchavadze Ave, Tbilisi 0179, Georgia*
[y]*Universidad de Los Andes, Apartado Aereo 4976, Carrera 1E, no. 18A 10, CO-Bogota, Colombia*
[z]*Dept. of Physics, Wisconsin University, Madison, WI 53706, United States*
[aa]*Vilnius University, Vilnius, Lithuania*

*E-mail:* kslee0421@korea.ac.kr

ABSTRACT: We report on a systematic study of double-gap and four-gap phenolic resistive plate chambers (RPCs) for the Phase-2 upgrade of the CMS muon system at high η. In the present study, we constructed real-sized double-gap and four-gap RPCs with gap thicknesses of 1.6 and 0.8 mm, respectively, with 2-mm-thick phenolic high-pressure-laminated (HPL) plates. We examined the prototype RPCs with cosmic rays and with 100-GeV muons provided by the SPS H4 beam line at CERN. To examine the rate capability of the prototype RPCs both at Korea University and at the CERN GIF++ facility, the chambers were irradiated with $^{137}$Cs sources providing maximum gamma rates of about 1.5 kHz cm$^{-2}$. In contrast to the case of the four-gap RPCs, we found the relatively high threshold on the produced detector charge was conducive to effectively suppressing the rapid increase of strip cluster sizes of muon hits with high voltage, especially when measuring the narrow-pitch strips. The gamma-induced currents drawn in the four-gap RPC were about one-fourth of those drawn in the double-gap RPC. The rate capabilities of both RPC types, proven through the present testing using gamma-ray sources, far exceeded the maximum rate expected in the new high-η endcap RPCs planned for future phase-II runs of the Large Hadron Collider (LHC).

**Contents**

**1. Introduction**

RPCs are a part of the Compact Muon Solenoid (CMS) [1-4], which plays as an important role in *e.g.* searches for new physics and in the recent discovery of the Higgs boson [5]. Inside the muon system, RPCs are mainly used as trigger devices, but also contribute to the reconstruction of muons. As illustrated in Fig. 1, the current CMS RPC system covering a pseudorapidity range of $|\eta| < 1.6$ comprises six trigger stations in the barrel and four trigger stations in the endcap sections of the CMS detector [6, 7]. The fourth RPC trigger station in the endcap section (the zone in the upper-right solid line of Fig. 1) has been recently installed to reinforce the muon-trigger efficiency of the CMS for the LHC runs, starting in 2015 [8, 9].

The current CMS endcap RPCs comprise 2-mm-thick double-gap chambers, each covering an azimuthal ($\varphi$) angular range of 10°. Each RPC chamber contains three sectors of 32 trapezoidal strips, each covering a detection range of 0.31° in $\varphi$ and 0.1 in $\eta$ [4].

The CMS RPC group recently addressed the need for an extension of the RPC system towards higher rapidity (the region in the bottom red box in Fig. 1) to maintain the present muon trigger performance during the future 14-TeV LHC runs with a maximum luminosity of about $5 \times 10^{34}$ cm$^{-2}$ s$^{-1}$ [11]. It is expected that these new RPCs (labeled as RE3/1 and RE4/1 in Fig. 1) will face a beam-related background with a maximum rate of about 600 Hz cm$^{-2}$ in the maximum $\eta$ division, an order of magnitude increase (by a factor at least 10 times higher) over the background rate previously experienced with the 8-TeV beam with a maximum luminosity of $6 \times 10^{33}$ cm$^{-2}$ s$^{-1}$ during past LHC operations [11, 12].

Improving the detector sensitivity represents the most relevant solution for proper RPC operations in an environment of a high-rate background. Furthermore, a lower-gain avalanche-mode operation should improve the longevity of the RPC detectors. In the present study, we examined two different types of phenolic RPCs: double-gap and four-gap RPCs, with single-gap thicknesses of 1.6 mm and 0.8 mm, respectively. Four real-sized prototype RPCs (two of each type) with 2-mm-thick phenolic HPL were constructed and tested with muons and gamma sources to probe the detector performances.

Previous reports have described the structure of the four-gap RPCs (each composed of two bi-gaps) [13, 14]. We measured the mean value of the bulk resistivity of the phenolic HPL panels used for the construction of the present prototype RPCs as about $3 \times 10^{10}$ Ω·cm at a temperature of 20°C. Figure 2 shows the layout of 128 readout strips in the prototype RPCs. We adopted the old RE1/1-type geometry for the prototype RPCs because its strip pitches and lengths approximately matched those envisioned for the future RE3/1 and RE4/1 RPCs.

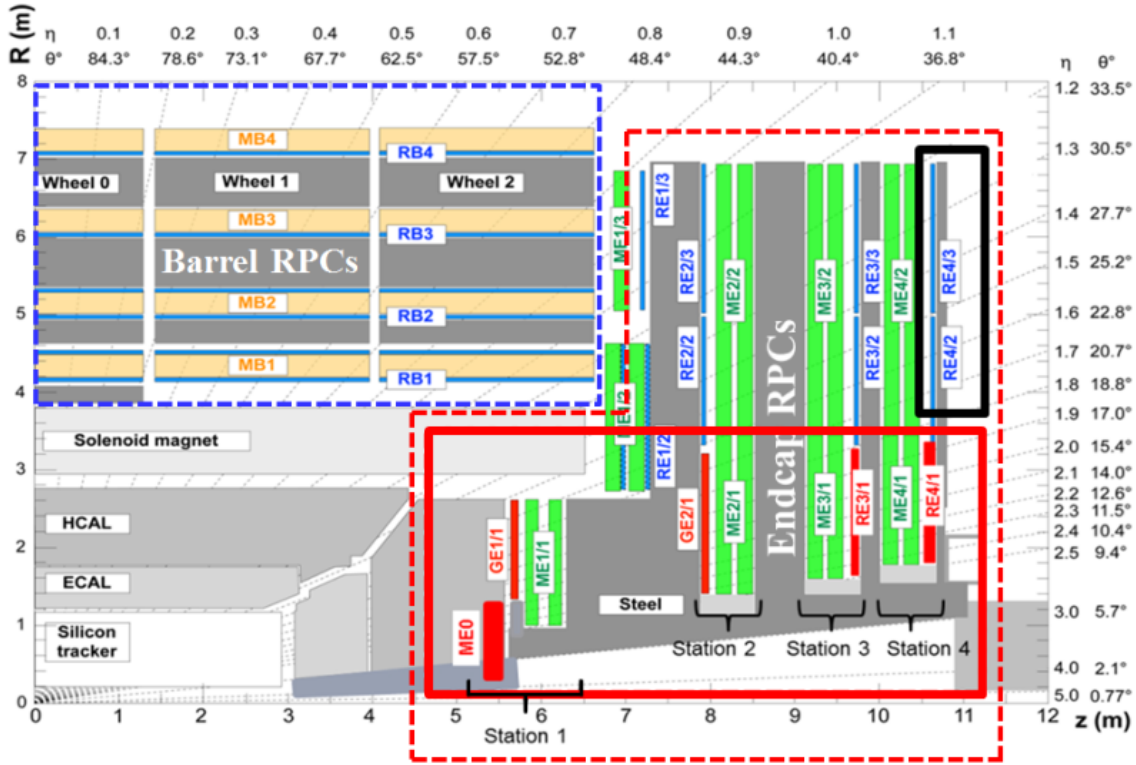

Fig. 1. Quadrant of the CMS detector, showing the muon system inside the dashed boxes, with the current RPCs in blue.

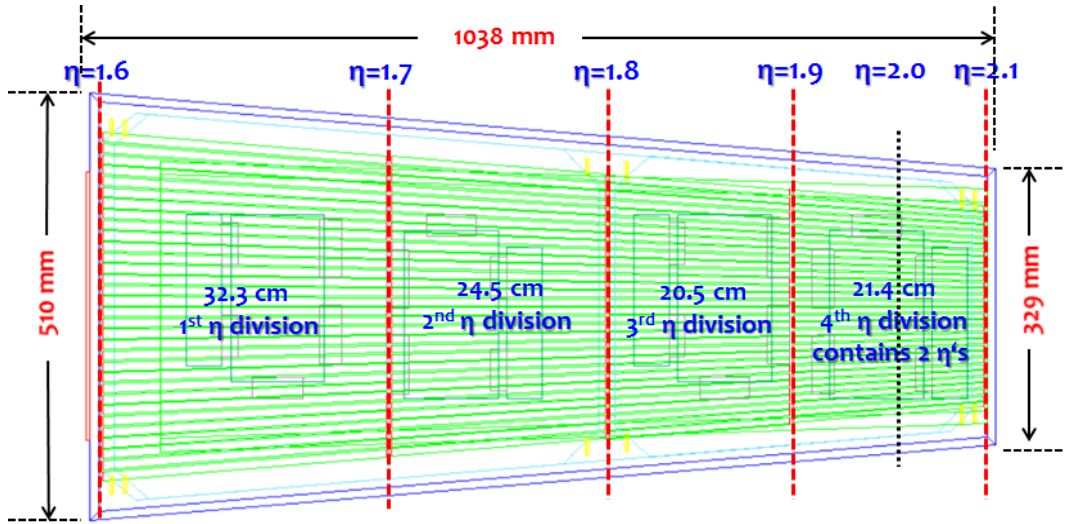

Fig. 2. Layout of 128 readout strips in the prototype RPCs constructed with the old RE1/1 type geometry.

## 2. Test facilities

We tested the prototype RPCs with cosmic rays at Korea University, and in parallel for 100 GeV muons provided by the CERN SPS H4 test beam line. To study the rate capability of the prototype RPCs, they were exposed to $^{137}$Cs gamma-ray sources, both at Korea University and at the CERN new Gamma Irradiation Facility (GIF++). The current activity levels of the sources in Korea and in the GIF++ were 5.55 GBq and 13.9 TBq respectively.

For the radiation test at Korea University, we placed each RPC type at a distance of 36 cm from the source, as shown on the left in Fig. 3. For the muon-beam test, we placed the two RPC types close together, installed at about 5 m from the GIF++ source, as shown on the right in Fig. 3. In both tests, the maximum rate of the gamma background on the prototype RPCs was 1.5 kHz cm$^{-2}$.

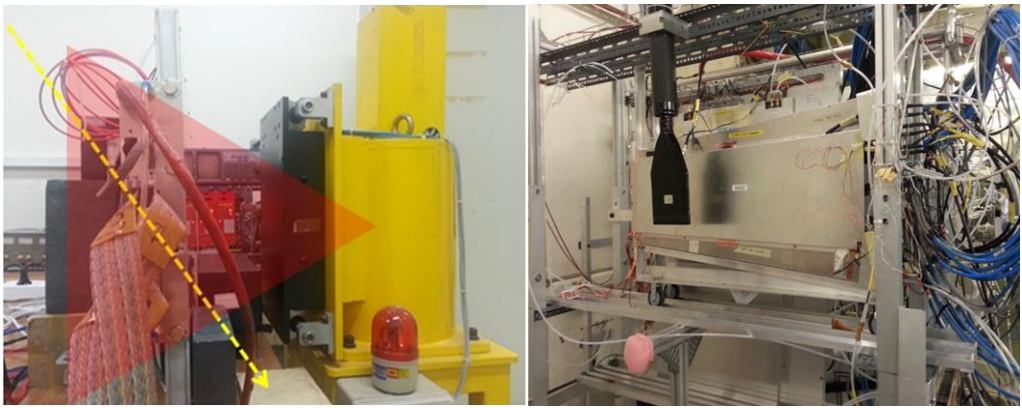

Fig. 3. Prototype RPCs installed at a distance of 36 cm from the 5.55-GBq $^{137}$Cs source at Korea University (left), and at about 5 m from the 13.9-TBq $^{137}$Cs source at the GIF++ (right).

The signals that arose in the RPC strips were transferred to front-end-electronics (FEE) via 50-Ω coaxial cables. We appropriately delayed the low-voltage differential signaling (LVDS) outputs of the FEEs, transferred to a multi-hit time-to-digital converter (TDC) via 34-pin twisted-pair cables. For the cosmic-muon test at Korea University, we used two different FEE types for the digitization of the signals: charge-sensitive FEEs for the operation of the current CMS RPCs; and voltage-sensitive FEEs developed specifically for the present study of RPCs. For the beam test at the GIF++, we used only CMS RPC FEEs for the digitization. The threshold values for the digitization in the CMS RPC FEEs ranged from 170 to 220 fC; those applied in the voltage-sensitive FEE ranged from 0.6 to 1.0 mV.

The gas mixture for the RPC operation used at both sites was composed of 95.2% $C_2H_2F_4$, 4.5% $i$-$C_4H_{10}$, and 0.3% $SF_6$. We added water vapor with a mass ratio of about 0.3% to the gas mixture to maintain a constant bulk resistivity for the HPL panels in the prototype RPCs.

## 3. Results

### 3.1 Threshold dependence

Figure 4 shows detection efficiencies ($\varepsilon_\mu$) and mean cluster sizes ($<C_s>$) as a function of high voltage for cosmic muons impinging on the first (left) and the third (right) $\eta$ division of the double-gap RPCs at Korea University. The mean values of the strip pitches for the first and the third $\eta$ divisions are 12.5 and 8.5 mm, respectively. We labeled the data obtained at the thesholds of 0.6 and 1.0 mV (voltage-sensitive FEEs) and of 180 fC (charge sensitive FEEs) by open and full circles and inverted triangles, respectively. We converted the applied high voltages to the effective values, $HV_{eff}$, under the standard conditions of $P = 1013$ hPa and $T = 293$ K [14]. As shown on the right in Fig. 4, $<C_s>$ of the muons measured at the narrower-pitch strips and at the lower threshold appeared especially high and increased more rapidly with $HV_{eff}$, while $\varepsilon_\mu$ appeared rather insensitive to the choice of the threshold.

We attributed the rapid increase of $<C_s>$ with $HV_{eff}$ in the data in Fig. 4 to the subservient hits induced by strong capacitive inductions among the narrow-pitch strips. Therefore, the relatively higher value of threshold appeared effective to suppress the magnitude of $<C_s>$. Nevertheless, fine adjustments of the threshold values in accordance with the strip pitches would be conducive to achieving a consistent hit structure over the whole detector surface.

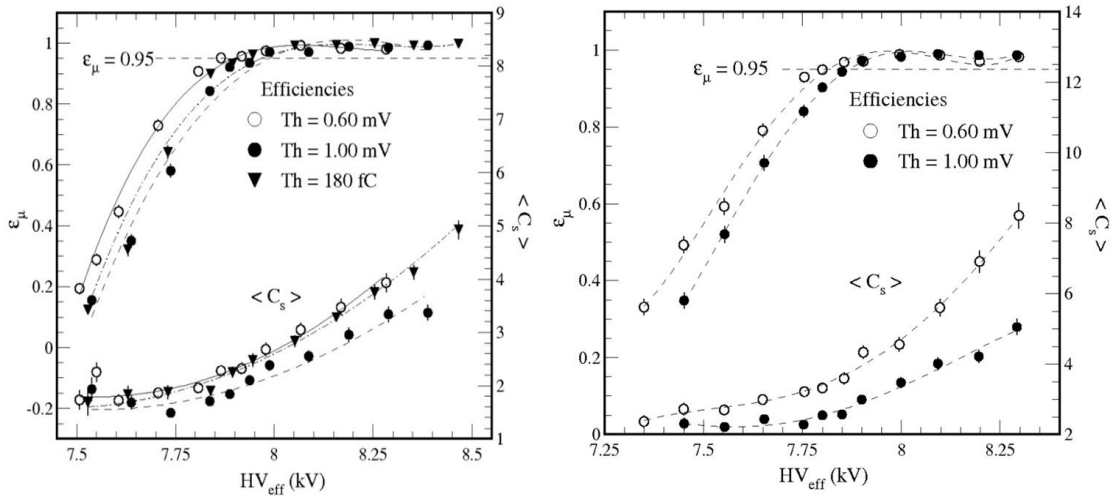

Fig. 4. $\varepsilon_\mu$ and $<C_s>$ for cosmic muons as a function of $HV_{eff}$ at Th = 0.6 (open circles) and 1.0 mV (full circles), and Th = 180 fC (inverted triangles), tested for the first (left) and the third (right) $\eta$ divisions of the double-gap RPCs at Korea University.

Figure 5 shows $\varepsilon_\mu$ and $<C_s>$ as a function of high voltage, measured with cosmic muons on the first (left) and the third (right) $\eta$ divisions of the four-gap RPCs, also at Korea University. We labeled the data obtained at the thesholds of 0.6, 0.75, and 1.0 mV by open circles, full circles, and squares, respectively.

In contrast to the case of the double-gap RPCs, both $\varepsilon_\mu$ and $<C_s>$ shift significantly with increasing the threshold value. The relatively larger sensitivity of $\varepsilon_\mu$ to the threshold for the four-gap RPCs implies an advantage, enhancing the detector sensitivity by lowering the threshold without unnecessarily magnifying the cluster sizes. Lowering the threshold (*i.e.*, enhancing the detector sensitivity) would be conducive to achieving higher rate capability, as well as to improving the longevity of the gas gaps.

## 3.2 Pickup charges

We linearly amplified the pickup signals from the cosmic muons in the two prototype RPCs by a factor 10, before sending them to two 400-MHz four-channel fresh analog-to-digital converters (ADCs). We carefully calibrated the amplitudes of the ADC values using well-defined square pulses provided by a pulse generator. We set the time window allowed for the integration of the pickup charges to 75 ns.

Figure 6 shows the mean pickup charges <$q_e$> (triangles, by ADCs) and $\varepsilon_\mu$ (by TDCs) as a function of HV$_{eff}$ measured at Th = 0.6 (open circles), 0.75 (full circles), and 1.0 mV (squares), for the first $\eta$ region of the double-gap (left) and the four-gap (right) RPCs. For the double-gap RPCs, the slope of the exponential growth of <$q_e$> decreased continuously with HV$_{eff}$. When $\varepsilon_\mu$ reached 0.95, the reduction of the slope became significant, allowing us to obtain the ample plateau size for the operation.

However, the exponential growth of the <$q_e$> measured with the four-gap RPCs appeared rather slow and monotonous, even at the low HV$_{eff}$. As the result, we expected that we could effectively shift the dynamic range of operation toward the lower HV$_{eff}$ by lowering the threshold, without losing the significant size of the operational plateau.

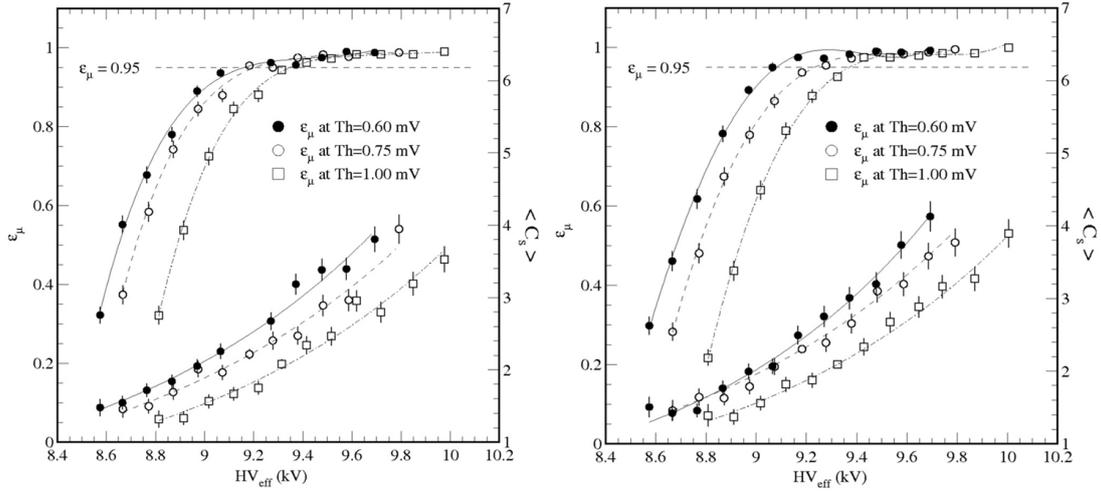

Fig. 5. $\varepsilon_\mu$ and <$C_s$> for cosmic muons as a function of HV$_{eff}$ at Th = 0.6 (open circles), 0.75 (full circles), and 1.0 mV (squares), tested for the first (left) and the third (right) $\eta$ divisions of the four-gap RPCs at Korea University.

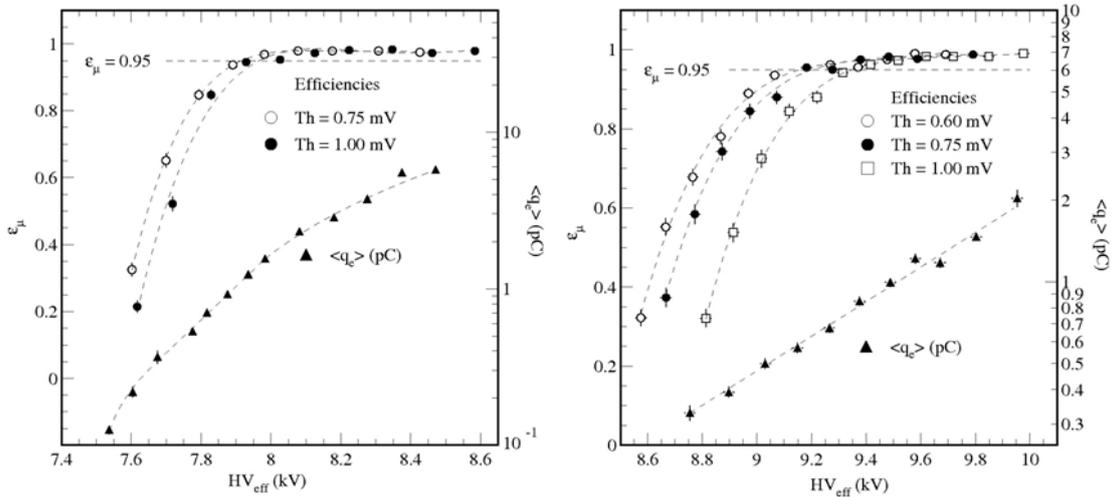

Fig. 6. <$q_e$> (triangles) and $\varepsilon_\mu$ measured at Th = 0.6 (open circles), 0.75 (full circles), 1.0 mV (squares), tested for the first $\eta$ region of the double-gap (left) and the four-gap (right) RPCs, as a function of HV$_{eff}$.

### 3.3 Influence of gamma background

We estimated the gamma-ray flux impinging on the RPCs placed at 36 cm from the 5.55 GBq $^{137}$Cs gamma source at Korea University to be 0.32 MHz cm$^{-2}$. Figure 7 shows $\varepsilon_\mu$ for the cosmic muons (top symbols scaled on the left axis) and the gamma rates $R_\gamma$ (bottom symbols scaled on the right axis) at Th = 0.6 (full circles), 0.75 (open circles), and 1.0 mV (squares), as a function of HV$_{eff}$, tested for the first $\eta$ divisions of the double-gap (left) and the four-gap (right) RPCs. We labeled the gamma-induced currents drawn in the RPCs as triangles. In the mid-section of efficiency plateau, the gamma-induced currents drawn in the four-gap RPCs were about one fourth of those drawn in the double-gap RPCs. Considering the high particle rates expected in the RE3/1 and RE4/1 RPCs, the smaller avalanche charges of the four-gap RPCs should provide a fair advantage to guarantee longevity.

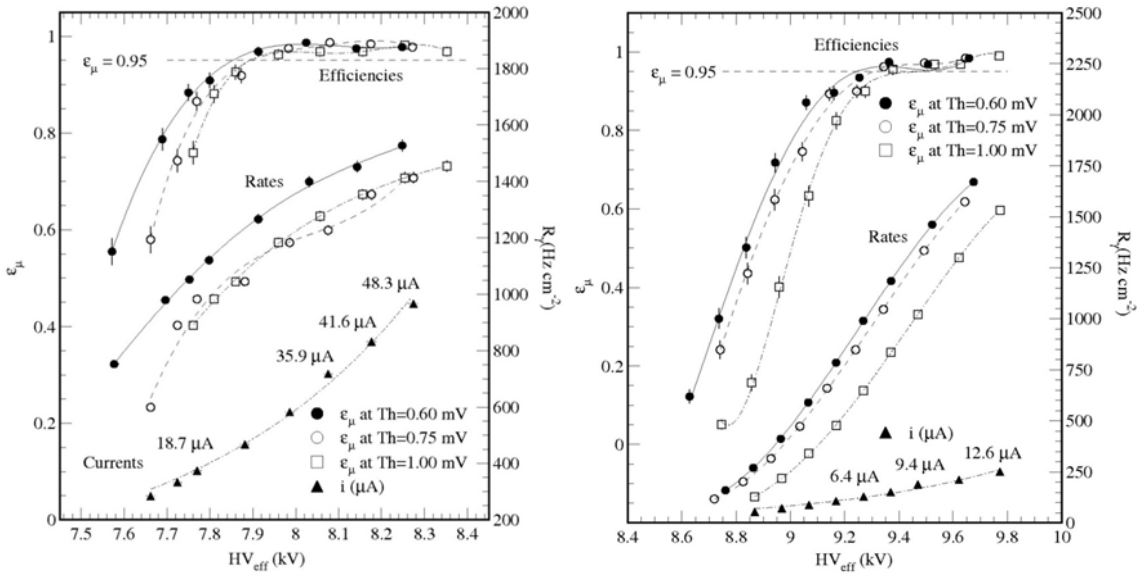

Fig. 7. $\varepsilon_\mu$ for the cosmic muons (top symbols scaled on the left axis) and the gamma rates $R_\gamma$ (bottom symbols scaled on the right axis) at Th = 0.6 (full circles), 0.75 (open circles), and 1.0 mV (squares), as a function of HV$_{eff}$, tested for the first $\eta$ section of the double-gap (left) and the four-gap (right) RPCs.

The influence of the gamma background on $\varepsilon_\mu$ and $<C_s>$ for the 100-GeV muons provided by the H4 beam line is shown in Figs. 8 and 9, respectively. For both detectors, we adjusted the threshold to 190 fC. We adjusted the intensity of the gamma rays emitted from the $^{137}$Cs gamma source at the GIF++ (impinging on the RPCs) by three-step attenuations using tungsten filters of various thicknesses. We measured the maximum gamma rates present with the attenuation factors of 69 (blue squares) and 2.2 (red circles) as 110 and 1545 Hz cm$^{-2}$ for the double-gap RPCs (placed relatively closer to the gamma source) and 72 and 1055 Hz cm$^{-2}$ for the four-gap RPCs, respectively.

As shown in Figs. 8 and 9, we observed fairly insignificant shifts in HV$_{eff}$ yielding $\varepsilon_\mu = 0.95$ (due to the presence of gamma rates of about 1 kHz cm$^{-2}$), while observing noticeable reductions in $<C_s>$. The low sensitivity of the efficiency to the gamma rate should serve to guarantee a reliable operating condition for the trigger RPCs in the high-rate background environment in future CMS experiments.

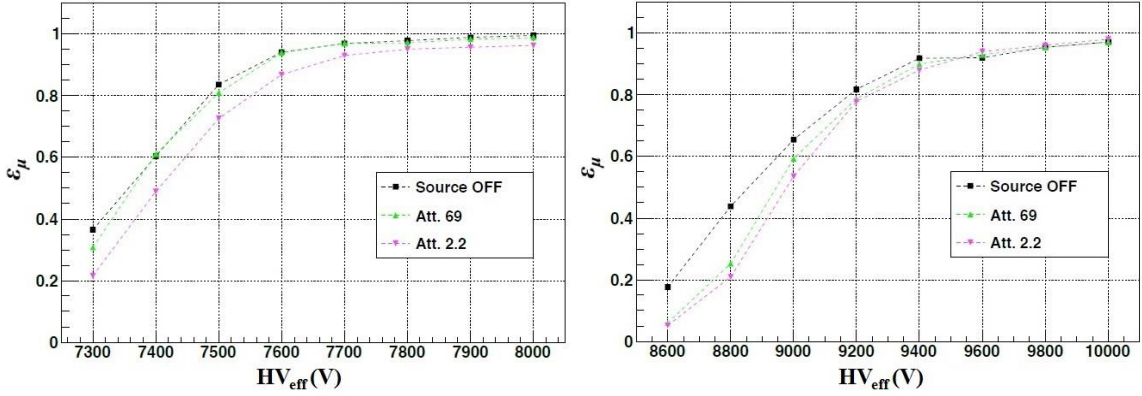

Fig. 8. $\varepsilon_\mu$ for the 100-GeV muons provided by the H4 beam line as a function of $HV_{eff}$, tested for the third $\eta$ divisions of the double-gap (left) and the four-gap RPCs (right). The text describes the details for the symbols.

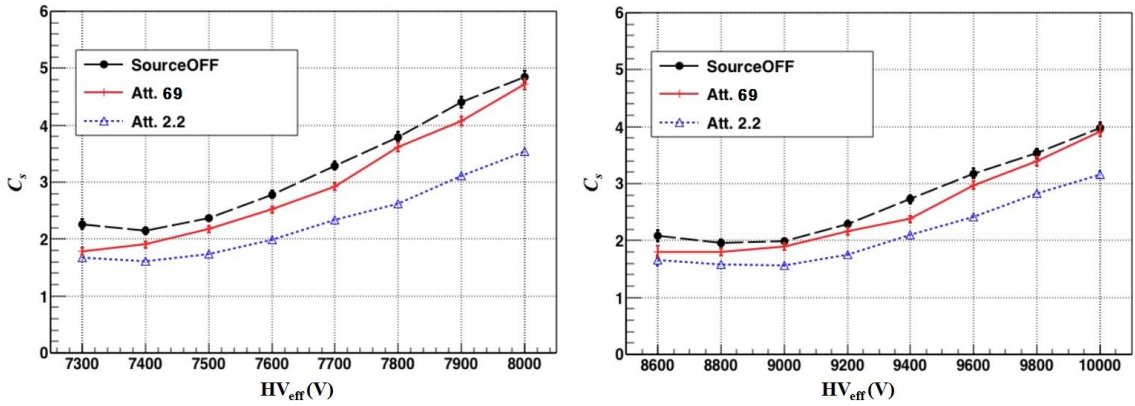

Fig. 9. $<C_s>$ for the 100-GeV muons provided by the H4 beam line as a function of $HV_{eff}$, tested for the third $\eta$ divisions of the double-gap (left) and the four-gap RPCs (right). The text describes the details for the symbols.

## 4. Conclusions

In the present research, we examined two different phenolic RPC types with cosmic rays at Korea University and 100 GeV muons provided by the SPS H4 beam line at CERN. We examined the rate capabilities of the prototype RPCs with gamma-ray hits of maximum 1.5 kHz cm$^{-2}$ provided by the $^{137}$Cs gamma sources at Korea University and the GIF++ irradiation facility at CERN. We drew the following conclusions from the present research:

(1) For the 1.6-mm double-gap RPC, a relatively high threshold value is preferred to suppress the magnitude of $<C_s>$ and the rapid increase with $HV_{eff}$. Fine adjustments of the threshold values in accordance with the strip pitches would be conducive to achieving a consistent hit structure over the whole detector surface.

(2) The relatively large sensitivity of $\varepsilon_\mu$ to the threshold value for the four-gap RPC implied an advantage, namely enhancing the detector sensitivity by lowering the threshold. In the mid-section of the efficiency plateau, the gamma-induced currents drawn in the four-gap RPC were about one fourth of those drawn in the double-gap RPC. Considering the high particle rates expected at the position of the future RE3/1 and RE4/1 RPCs, the smaller avalanche charges of the four-gap RPCs appeared fairly advantageous to guarantee the longevity of the trigger RPCs.

(3) For both RPC types, we observed quite insignificant shifts of the $HV_{eff}$ curves due to the presence of the gamma rates of about 1 kHz cm$^{-2}$. The low sensitivity of the efficiencies to the gamma rate should guarantee a reliable operation of the trigger RPCs in the high-rate background environment in future CMS experiments.

## Acknowledgments

This study was supported by the National Research Foundation of Korea (grant number NRF-2013R1A1A2060257). Most of all, we would like to give our special thanks to all team members dedicated the beam test and to the CERN EN and EP departments for the facility infrastructure support.

## References

[1] P. Paolucci *et al.*, J. Instrument **8** P04005 (2013).

[2] The CMS Collaboration, J. Instrument **5** T03017 (2010).

[3] The CMS Collaboration, *Detector Performance and Software, Technical Design Report* **Vol. I**, CERN/LHCC 2006-01, February 2006.

[4] CMS Collaboration, *The Muon Project, Technical Design Report*, CERN/LHCC 97-32, December 1997.

[5] CMS Collaboration, Phys. Lett. B **716** 30 (2012).

[6] M. Konecki on behalf of the CMS Collaboration, J. Instrument **9** C07002 (2014).

[7] S. Costantini on behalf of the CMS Collaboration, J. Instrument **8** P03017 (2012).

[8] M. Tytgat on behalf of the CMS Collaboration, J. Instrument **8** T02002 (2013).

[9] S. K. Park for the CMS Collaboration, J. Instrument **7** P11013 (2013).

[10] '*Phase-2 Technical Proposal*', CERN-LHCC-2015-010; LHCC-P-008.

[11] '*CMS RPC muon detector performance with 2010-2012 LHC data*', G. M. Pugliese on behalf of the CMS Collaboration, XII Workshop on Resistive Plate Chambers and Related Detector, Beijing February 2014.

[12] '*Radiation Background with the RPCs at the CMS Experiment*', S. Costantini on behalf of the CMS Collaboration, XII Workshop on Resistive Plate Chambers and Related Detector, Beijing February 2014.

[13] K. S. Lee on behalf of the CMS Collaboration, J. Instrument **9** C08001 (2014).

[14] S. K. Park *et al.*, Nucl. Instr. Meth. A **680**, 134 (2012).